# Electrochemically controlled polymeric device: a memristor (and more) found two years ago

Victor Erokhin[1,2] & M.P. Fontana[1]

[1]Department of Physics, University of Parma and CRS SOFT CNR-INFM, Viale Usberti 7A, Parma (PR) 43100 Italy

[2]Institute of Crystallography, Russian Academy of Sciences, Leninsky pr. 59, Moscow, 119333 Russia

**In the last few decades, the age old dream of building an artificial brain, i.e. using biological cognitive systems as a benchmark and inspiration to fabricate complex material assemblies which can learn, make decisions, analyse information in a highly parallel way: in other words, highly efficient bio-inspired information processors, has acquired the concreteness of real life research. Many programs and projects in materials science, nanotechnologies, ICT, bio-sciences use the relevant biological systems and processes as the basic paradigm for the research. However the enormous complexity of even the simplest brains still puts a barrier to the realization of such ambitions. Hence most of the current research (apart from theoretical modelling and simulations) deals with the fabrication and characterization of specific components which are expected to mimic in their functional behaviour neurons, synapses, or use biological molecules to build innovative sensing or electronic components[1-5]. Recently a non-linear inorganic thin film two electrode device has been reported[6] which the Authors claim to be the very first memristor, i.e. a resistor with memory or which can learn. The**

**device was based on a spatially graded ion doped TiO$_2$ thin (12nm) film. In this Letter we wish to show that a totally different "memristor" with even better characteristics (learning, memory) was reported in the literature since 2005[7-9] and discuss its properties and functioning principles, in the framework of using it as a main building block in complex functional networks for bio-inspired information processing.**

In the general framework of circuit theory, Chua had predicted the necessity of the existence of the fourth passive two-electrode electrical element establishing the connection between the magnetic flux and charge[10]. The element, called "memristor", should act according to the following relationship:

*dφ = Mdq*

where *dφ* and *dq* are variations of the magnetic flux and charge, respectively, while *M* is the memristance. Of particular interest are the non-linear characteristics of the element, for instance when *M* is not a constant but it is a function of the charge flow and elapsed time. In this case, the element properties may be influenced its functioning history: i.e. it will have memory. Apart from the obvious value of such device taken singly, here we wish to emphasize its potential as a key element of networks with adaptive properties, capable of learning, information analysis and decision making. In other words, its behaviour could mimic the features of synapses in the nervous system. Variation of the element properties (resistance, in particular) according to the previous experience or to external stimuli (e.g. appropriate electrical pulses) will play the same role as the variation of the synaptic weight functions in nervous system, leading to learning and memory mechanisms described by the Hebbian rule[11] (we do not consider here non-Hebbian algorithms of learning).

The recently reported memristor[6] has exhibited some variation of the element properties (resistance, in particular) according to the history of the voltage application. The reported results are rather interesting as they allow to describe some phenomena, previously observed in nanostructures[12-16], from the new view point. However, there are two limits to consider the element as a fundamental finding from basic and applied points of view. Firstly, it seems not completely correct to call the reported element as "memristor". As mentioned by the Authors, there is not a direct relationship between the variation of the magnetic flux and charge. Moreover, such direct relationship cannot be found in two-electrode devices, at least if based on the suggested physical principles. In fact, the element functioning was based on the longitudinal displacement of the doping impurities in the applied electric field, while the current variation was measured in the same direction. Instead, in order to obtain the connection between charge and magnetic flux, motion of charges in two perpendicular directions is required. Secondly, the reported experimental characteristics can be attributed also to a bistable electrical device rather then to an element which is purported to be used as a synaptic analogue, which must vary its resistance gradually according to its previous history.

In this respect, even if it was not specifically considered as "memristor", an electrochemically controlled hybrid ionic-conducting polymeric device[7], besides already having the memory and learning characteristics proposed for the $TiO_2$ device, has much more potential for application to bio-inspired information processing and in particular to the fabrication of complex adaptive networks, in which it would be the "synaptic" node[8-9].

The working principle of the element is based on the dramatic variation of the electronic conductivity in a thin (50nm) conducting polymer (polyaniline, PANI) multilayer in oxidized and reduced states[17]. Such variation is induced and regulated by ionic flux into (and out of) the PANI multilayer at the junction with a film of a solid electrolyte (Li-doped Polyethylene oxide, PEO)[18].

The scheme of the element is shown in Fig. 1. Although when connected in the circuit, the element would have two working electrodes only, the functioning of the single element is best described by considering a third "reference" (i.e. kept at ground potential) electrode. Thus there are two currents flows (electronic and ionic) which can be measured. The position of the second ammeter is also shown in Fig. 1 by a dashed line.

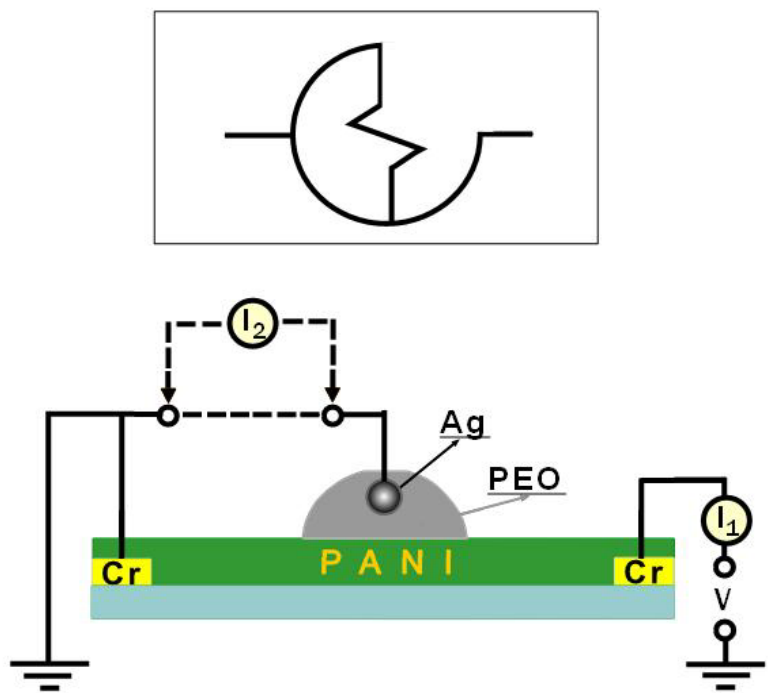

Fig. 1. Scheme of the electrochemical element and its symbol (upper panel) for electric circuits. Active layer was formed from the conducting polymer (PANI) with attached two metal electrodes. A stripe of solid electrolyte (lithium perchloride doped polyethylene oxide (PEO)) was formed in the central part of the PANI layer in order to provide a suitable medium for redox reactions. The area of PANI layer under the electrolyte is the active zone. The reference potential was provided by a silver wire inserted into the electrolyte. The wire was connected to the one of two metal electrodes as shown in the figure.

The mechanisms of resistance variation occurring within the element are illustrated in Fig. 2.

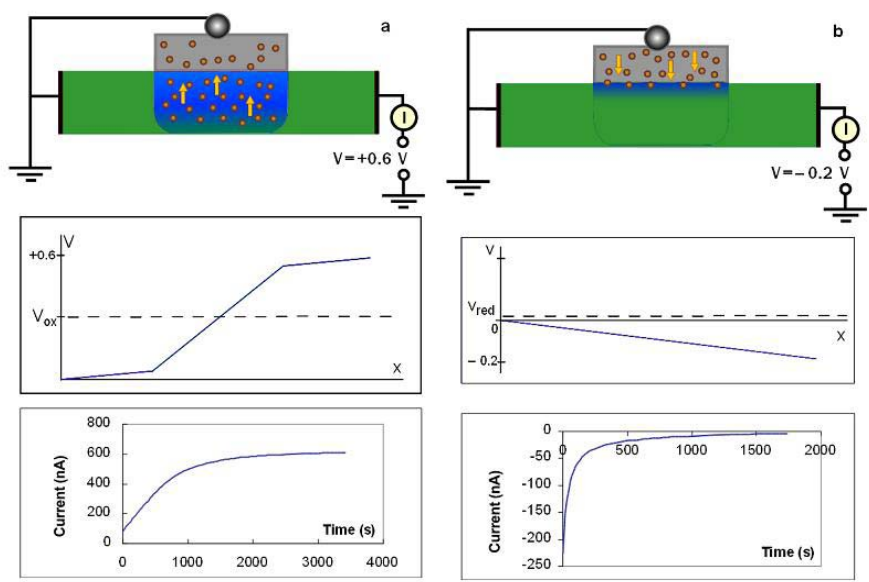

Fig. 2. Mechanisms of conductivity variation in the electrochemical polymeric element. Green areas correspond to the PANI in its oxidized conducting state, while blue areas represents PANI in reduced insulating form. The orange dots represent the Li ions, the arrows the prevalent direction of motion. (a) Positive potential is applied to the initially insulating (blue area) device. Electrical potential profile along the PANI layer is shown in the central part. Only some part of the active zone is at the potential higher than the oxidizing potential. Transformation of the PANI into the conducting state will take place only in this restricted zone. However, the transformation will result in the redistribution of the potential profile and new zones (shifted into left direction) will arrive to the oxidizing potential and will have the possibility to be transferred into the conducting state. Therefore, in this case we will have gradual displacement of the conducting zone boundary, that determines much slower kinetics of the conductivity variation (bottom part) for the positive voltage. (b) Negative voltage is applied to the initially conducting element. Electrical potential profile along the PANI layer is shown in the central part. All active zone is under the reduction potential. Therefore, reduction and transformation of PANI into the insulating state takes place simultaneously in the whole active area providing rather fast kinetics of the transformation (bottom part).

The element is characterized by two charge flows in perpendicular directions. Thus, this electrochemical device is more similar to the hypothetical memristor, suggested 40 years ago by Chua[10]. Ionic flow is determined by the actual potential of the active area with respect to the reference value (ground potential in this case) and provides incoming-outgoing Li$^+$ ions to the PANI layer, thereby varying its conductivity[18]. The actual resistance of the active zone is determined by the time integral of the ionic current (transferred charge). It is important that the value of the ionic current is more then one order of magnitude less than the current in PANI layer. Therefore, the actual measured current of the element in the conducting state, i.e. the sum of the electron and ionic currents, is mainly determined by the first contribution.

This configuration allows the element to memorize the information on its previous experience in signal transmitting. Let us consider the current temporal behaviour for the positive bias (Fig. 2a bottom panel). It exhibits a rather slow increase of the conductivity. This establishes the basis for unsupervised learning and synaptic-like behaviour. In fact, if we consider a network constructed from many such elements, connected in a complex way in order to provide numerous different pathways for the signal between any input-output pairs, there will be an increase of the conductivity of the those elements which are involved into the most frequently used pathways. Thus, similar stimuli (signals configuration on input electrodes), repeated in the future, will have an increased probability to result in similar conclusions (output signals). This is very similar to synaptic function (learning) in real biological systems, summarized in the Hebbian rule.

The behaviour of the element at negative bias is also interesting. It can be used to avoid network saturation: periodic short-time application of negative signals between all input-output pairs will prevent reaching the highest conducting level for all elements of the network, which would preclude further learning. In addition, this characteristic allows in principle supervised learning for networks of such elements. Application of negative bias between selected input-output pairs will destroy *a priori* wrong but statistically preferential signal pathways, established occasionally during unsupervised learning.

At the level of simple circuits and networks, we have already demonstrated the validity of the above considerations. A circuit based on one electrochemical element only, has demonstrated unsupervised learning[19]. A simple network of 8 interconnected electrochemical elements, has shown supervised learning, i.e. a variation of the output signal according to the external training procedure at the input electrodes[8].

Another, unexpected, interesting feature of the element was observed when the material of the reference electrode was changed[20]. Simple substitution of the silver wire with a graphite stripe resulted in the appearance of current auto-oscillations at constant applied voltage **in a simple two-electrode element!** A typical curve is shown in Fig. 3. The observed phenomenon is connected to charge accumulation by the material of the reference electrode (graphite). Thus, the reference potential is not fixed anymore. Several processes must be considered to explain such oscillating behaviour of the current: redox reactions in the active area of PANI layer according to its actual potential with respect to reference point; redistribution of the applied voltage profile along the PANI layer length according to the conductivity state of its different parts; accumulation and release of the charges in the material of the reference electrode resulting in the variation of its potential. A heuristic model[21], based on the consideration of all possible electrical phenomena in the element, has resulted in qualitative explanation of the observed characteristics, allowing the connection between electrical behaviour and materials properties, such as ionic diffusion and redox potentials. However, on a more fundamental level, it is possible to see several similarities of the observed phenomenon with well-known Belousov-Zhabotinsky (BZ) reaction[22]. In both cases we deal with reduction and oxidation processes, while accumulation and release of the charge at the reference point together with cyclic redistribution of the applied voltage profile in our device can be considered as analogues of the processes responsible for the production and inhibition of the catalyser in BZ reactions. If so, this would be the first demonstration of the appearance of rhythmic oscillations of electrical signals, instead of optical or viscoelastic properties variation cycles, usually observed in such reactions[23]. This feature is very important as similar non-equilibrium processes occur in biological systems. More specifically, the behaviour of our oscillating device seems to mimic that of a specific neuron in that part of the *Limnea Stagnalis* brain devoted to feeding behaviour[24].

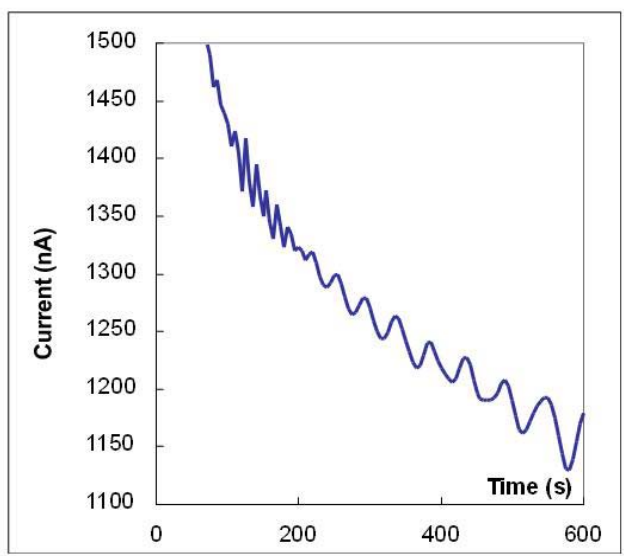

Fig. 3. Temporal dependence of the current of the electrochemical device at constant applied voltage of +2.0 V with a graphite reference electrode which allows the charge accumulation.

In conclusion, here we have shown that a "memristor" was already found more than two years ago, and that its characteristics make it much more promising as a basic building block of simple or complex adaptive electrical circuits, with all the advantages of the "bottom up" structured assembly fabrication procedures of molecular systems. The memory and learning properties of the device alone have already been demonstrated, and the outlook for application to more complex functional networks (deterministic or self-assembled) seems very promising. Statistically formed hybrid networks of conducting-ionic polymer fibers[25] have also shown similar features to deterministic elements, indicating that the bottom-up statistical self-assembly approach to the fabrication of complex adaptive networks is a viable alternative to standard fabrication.

Furthermore, although most of our work was performed on macroscopic systems (apart from the nanoscale of the active conducting polymer film thickness), our preliminary results show that there is in principle no obstacle to miniaturization to the micron or the submicron scale, yielding organic signal processors with high parallelism and a density of active components comparable to that of standard inorganic microelectronics.

Finally, such systems, being, as they are, based on basic cognitive processes of biological systems, have the full potential of serving as synthetic testing grounds for learning theories and algorithms.

**Acknowledgements** We acknowledge the financial support of the Future and Emerging Technologies (FET) programme within the Seventh Framework Programme for Research of the European Commission, under the FET-Open grant agreement BION, number 213219. We wish to thank Prof. A. Schüz for inspiring information.